\documentclass[11pt]{article}

\usepackage{geometry}                
\usepackage{graphicx}
\usepackage{amssymb}
\usepackage{epstopdf}
\usepackage{fancybox}
\usepackage{color}
\usepackage{verbatim}
\usepackage{url}

\usepackage{authblk}

\usepackage{enumerate}

\newcommand{\Aut}{{\rm Aut}}

\newcommand{\Ker}{{\rm Ker}}

\newcommand{\Sym}{{\rm Sym}}
\newcommand{\Sol}{{\rm Rad}}

\newcommand{\Z}{{\rm Z}}
\newcommand{\1}{{\bf 1}}

\newcommand{\nrm}[2]{{#1}_{#2}}

\newtheorem{theorem}{Theorem}[section]

\newtheorem{lemma}[theorem]{Lemma}
\newtheorem{proposition}[theorem]{Proposition}
\newtheorem{remark}[theorem]{Remark}

\newcommand{\calA}{{\mathcal A}}
\newcommand{\calB}{{\mathcal B}}

\newcommand{\calL}{{\mathcal L}}
\newcommand{\calM}{{\mathcal M}}

\newcommand{\fra}{{\mathfrak a}}
\newcommand{\frb}{{\mathfrak b}}

\newcommand\restr[2]{{
  \left.\kern-\nulldelimiterspace 
  #1 
\vphantom{\big|} 
  \right|_{#2} 
  }}

\newcommand\InnH[1]{{\iota(#1)}}

\newcommand\inn[1]{\iota(#1)}

\def\C{{\rm C}}
\def\CH{{H \C }}

\newcommand{\normal}{\vartriangleleft}
\newcommand{\normaleq}{\trianglelefteq}

\newcommand{\nin}{\noindent}
\newcommand{\mni}{\medskip\noindent}
\newcommand{\ms}{\medskip}
\newcommand{\bs}{\bigskip}
\newcommand{\bni}{\bigskip\noindent}
\newcommand{\ssk}{\smallskip}
\newcommand{\skni}{\smallskip\noindent}

\newcommand{\QED}{{\hfill $\Box$}}

\newcommand{\nice}{almost-solvable}
\newcommand{\niceness}{almost-solvability}
\newcommand{\Nice}{Almost-solvable}

\def\CSAuto{{\sc Comp\-Series\-Auto}}
\def\CSIso{{\sc Comp\-Series\-Iso}}
\def\AutLift{{\sc AutLifting}}

\newcommand{\PROBLEM}[1]
{{
\noindent
\hspace*{.5em}{{
\parbox[t]{5.3in}{
{#1}
}}
}
}
}

\newcommand{\COMMENT}[1]
{{
\Ovalbox{\begin{minipage}{5.35in}\ms\centerline{\begin{minipage}{5.1in}
#1
\end{minipage}\smallskip
}\end{minipage}}
}}

\title{Group Isomorphism with Fixed Subnormal Chains}

\author{Eugene M. Luks}

\affil{{Computer and Information Science Department}
\\\vspace*{.2em}{University of Oregon}}

\affil{email: {\tt {luks@cs.uoregon.edu}}}

\date{October 31, 2015}

\begin{document}

\maketitle

\begin{abstract}
In recent work, Rosenbaum and Wagner
showed that isomorphism of explicitly listed $p$-groups of order $n$ could
be tested in $n^{\frac{1}{2}\log_p n + O(p)}$ time, roughly
a square root of the classical bound.   The $O(p)$ term is entirely due
to an $n^{O(p)}$ cost of testing 
for isomorphisms that match fixed composition series
in the two groups.
We focus here on the fixed-composition-series
subproblem and exhibit a polynomial-time
algorithm that is valid for general groups.   
A subsequent paper
will construct canonical forms within the same time bound.
\end{abstract}

\section{Introduction}
\label{intro}

\ms
The complexity of testing isomorphism of groups of order $n$ (input via Cayley tables)  has long been
cited as $\,n^{\log_p n  + O(1)}$, 
where $p$ is the smallest prime divisor of $n$; this follows
immediately from the fact that the group requires no more than $\,\log_p n\,$ generators
\cite{LSZ77}.    Wagner \cite{Wa11} suggested that this might
be improved by a careful consideration of the isomorphisms that
match fixed composition series.     
While composition series have long been a staple for such computational problems
(e.g., \cite{FN67}), Wagner's
insight was that this could lead to an analyzable advantage
in terms of a provable worst-case bound.
Indeed, using that approach,
Rosenbaum and Wagner \cite{RW15} were able to improve the
bound for $p$-groups to $n^{(1/2)\log_p n + O(p)}$.   The $O(p)$ 
term is contributed by their 
$n^{O(p)}$ 
complexity analysis of
fixed-composition-series isomorphism in $p$-groups.
Their main result then uses the fact that there are $n^{(1/2)\log_p n + O(1)}$
composition series to consider.
Rosenbaum \cite{Ros13} extended the result to
solvable groups, achieving an $n^{(1/2)\log_p n + O(\log n/\log\log n)}$ time
for isomorphism, the fixed-composition series subproblem contributing to the
$\,\log n/\log\log n\,$ term.
Making use of a canonical-form version of the
fixed-composition-series subproblem,
Rosenbaum \cite{RosarX13} subsequently showed that a ``collision'' method
yields a square-root improvement.   That  result is striking and there
is much to be appreciated in the methods.   However, 
the composition-series-isomorphism subproblem remains appealing in its own right
and it appears to be susceptible to established algebraic and computational
methods.

\ms

Thus, we focus on the following problem.

\bs

\PROBLEM{
\CSIso\\[.5ex]
\hspace*{1.5 em} {\sc Given:} Groups $G_1,G_2$ given by Cayley tables;\\
\hspace*{5.25em} composition series\\
\hspace*{7.25em}$G_1=G_{1,0} \vartriangleright G_{1,1}  \vartriangleright G_{1,2}  \vartriangleright \cdots \vartriangleright G_{1,m}  =\1$, \\[.5ex]
\hspace*{7.25em}$G_2=G_{2,0} \vartriangleright G_{2,1}  \vartriangleright G_{2,2}  \vartriangleright \cdots \vartriangleright G_{2,m}  =\1$.\\[.5ex]
\hspace*{1em} {\sc Question:}
Is there an isomorphism $~f: G_1 \rightarrow G_2~$  such that \\[.5ex]
\hspace*{7.25em}
$f(G_{1,i})= G_{2,i}$, for $0\le i\le m$?
}

\bs

The treatments of \CSIso~for nilpotent and solvable groups
in \cite{Wa11}, \cite{RW15}, \cite{Ros13}, put great effort into reductions to instances of bounded-valence graph isomorphism so as to
plug in the main result of \cite{Luk82}.     However, underlying the latter was a method for set-stabilizers in permutation groups
which we can apply directly in a natural approach to \CSIso.    This results in both a better bound
and an extension to general groups.
Specifically,

\ssk

\begin{theorem}
\label{mainprob}
{\sl \CSIso~is in polynomial time.}
\end{theorem}

If one separates the elements needed for just the solvable-group case, the proof
can be compressed to a few lines.

\ms

It should be no surprise that the method actually returns {\it all\/} isomorphisms
in the form of a coset of the analogous automorphism group.   In fact,
our discussion concentrates mostly on finding automorphism groups.    
\bs

\PROBLEM{
\CSAuto\\[.5ex]
\hspace*{1em} {\sc Given:} A group $G$ given by Cayley table;\\
\hspace*{4.75em} a composition series
$G=G_0 \vartriangleright G_1  \vartriangleright G_2  \vartriangleright \cdots \vartriangleright G_m =\1.
 $\\[.5ex]
\hspace*{1em} {\sc Find:}
(Generators for) the group $\{f\in \Aut(G) \mid f(G_{i})= G_{i} \mbox{ for } 0\le i\le m\}$.
}

\bni
We prove


\begin{theorem}
\label{mainprobauto}
{\sl \CSAuto~is in polynomial time.}
\end{theorem}

\noindent
The application
to isomorphism then follows a quick and standard path.    

\ssk
\ms
An  immediate consequence of Theorem~\ref{mainprob} is  the time bound
$\,n^{(1/2)\log_p n + O(1)}\,$ for testing isomorphism
of groups of order $n$,
where $p$ is the smallest prime divisor of $n$;
this is due to the upper bound of 
$\, n^{(1+\log_p n)/2}\,$
on  the number of composition series (credited to Babai in \cite[Lemma 3.1]{RW15}).
Rosenbaum \cite{RosarX13}
has already realized that timing for isomorphism
using ``bidirectional collision'', though his method comes
at a substantial cost in space.    
Collision and other innovative methods in
\cite{RosarX13} also mesh well with our results, with
further implications for group isomorphism.
We will be pleased to borrow from those methods
in a follow-up paper which will first extend our \CSIso~algorithm
to the computation of  canonical forms.

\ms
We remark that Babai \cite{Ba12} has found a holomorph approach 
to \CSIso~which also improves the earlier bounds and is in polynomial time for solvable groups.

\bs
Our method for \CSAuto~involves repeated consideration
of a classic issue in automorphism-group computation.
Specifically, for $H\normaleq G$,  we are given $\calA \le \Aut(G/H)$ and $\calB\le \Aut(H)$,
and we want to determine the pairs $(\alpha,\beta)\in \Aut(G/H)\times \Aut(H)$
that ``lift'' to automorphisms of $G$, if any such exist.    The obstruction to such lifting
is easy to formulate algebraically and, in general, it poses a difficult
computational problem.   However, for the $\calA$ and $\calB$
that arise herein, we are able to view the obstruction in stages, 
each of which is resolvable using methods that are in polynomial time
for  solvable permutation groups.
We give two procedures for this.  An elementary method, described in \S\ref{firstL},
is all one needs for the resolution of
\CSAuto~for solvable groups, but it is not effective for
general groups.     The method is then strengthened in \S\ref{secondL} to one that
is
generally applicable.

In \S\ref{nice}, we recall the divide-and-conquer method for set-stabilizer
in permutation groups and show that it is in polynomial time for
the groups we encounter since they are shown to have solvable subgroups
of ``small'' index.

There are  two demonstrations of Theorem~\ref{mainprobauto}
in  \S\ref{auto}.
They exhibit two ways of breaking the problem down into instances of the
``lift'' scenario that are amenable to a set-stabilizer approach
(assuming they are not already in polynomial time via brute-force enumeration).

Theorem~\ref{mainprob} is succinctly resolved in
\S\ref{Iso} by viewing it
as an extended application of the methods of \S\ref{auto}.

\bni
{\bf Notation.}

\ms

Suppose $G$ is a group acting on the set $\Omega$.  
For $g \in G$, we denote the image of $\alpha \in \Omega$ under the
action of $g$ by $\alpha^g$.   For $\Delta\subseteq\Omega$, let
$
{G}_{\Delta} = \{g\in G \mid \Delta^g = \Delta\}
$.

The automorphism group of $G$ is denoted by $\Aut(G)$, which we view as
a subgroup of $\Sym(G)$.   Thus, for $H\le G$, $\Aut(G)_H$ denotes the group
of automorphisms stabilizing (or normalizing) $H$.


Implicit in the statement that $G$ is given by a Cayley table is the assumption that
the elements of $G$ can be listed in polynomial time.  
Permutation groups
are assumed to be input or output via generating sets.   A coset of
a permutation group $J$ is input or output via generators for $J$ and
a single representative.

For other concepts and notation, we
refer the reader to \cite{DM96}.   For background on
polynomial-time computability in
permutation groups, see \cite{KL90}, \cite{Luk82}, \cite{LM11}, \cite{Se03}.

\bs
\section{The key subproblem}
\label{key}


\ms

Throughout this section, we assume $G$ is given by a Cayley table and $H\normaleq G$.
Then   $\Aut(G)_H$ can be viewed as a subgroup of the
wreath product $\Sym(H) \wr \Sym(G/H)$, and especially a subgroup of
$\Sym(H) \wr \Sym(G/H)_{\{H/H\}}$.  
(The subscript $\{H/H\}$ signifies that we fix this single ``point''
in the permutation domain $G/H$.)~
For the reader's convenience
we give a more explicit indication of
the latter group, namely,
$$\Sym(G)_{H,G/H} := \{ \gamma\in\Sym(G)_H \mid
 \gamma \mbox{ permutes the cosets of }H\}.$$

\noindent
There is a natural homomorphism
$$
\Theta_{G,H}:     \Sym(G)_{H,G/H} \,\rightarrow\, \Sym(G/H)\times\Sym(H).
$$
\noindent
Inasmuch as $H\normaleq G$ will always be clear in context, we let $\Theta:=\Theta_{G,H}$.  For $x\in G$,
we let $\inn{x}$ denote the restriction to $H$ of the inner automorphism due on $x$, i.e., for $h\in H$ 
$h^{\inn{x}} =  x^{-1} h x$; this is extended to $X\subseteq G$ by $\InnH{X} = \{ \InnH{x} \mid x\in X\}$.
 Because of repeated usage, it is useful to
set $\C :=\{g\in G \mid \forall h\in H, gh=hg\}$, i.e., the centralizer of
$H$ in $G$.


\ms
\subsection{Lifting automorphisms from $G/H$ and $H$}
\label{autlifting}

Noting that
$\Theta(\Aut(G)_H)\,\le \Aut(G/H)\times\Aut(H)
$, we are concerned with the following problem.

\PROBLEM{
\medskip
{\sc AutLifting}\\[1ex]
\hspace*{1em} {\sc Given:} $H\normaleq G$;~
 $\calA\le \Aut(G/H)$;~ $\calB \le \Aut(H)$.\\[1ex]
\hspace*{1em} {\sc Find:}
$\,\Aut(G)_H \cap \Theta^{-1}(\calA\times \calB).\,$
\\[-.75ex]
}

\ms
There are a couple of natural ways to reduce Theorem~\ref{mainprobauto} to
polynomial-time instances of {{\sc AutLifting}.  We will describe these in  \S\ref{bottomup},\ref{topdown}.

\bs

The reader may already recognize {\sc AutLifting}
as a frequent issue in studies of group automorphism/isomorphism, theoretical or computational
(see, e.g., \cite[\S8.9]{HEO}).
Special cases are attacked with varied machinery and success.  
Our method is guided by properties of the relevant groups that enable polynomial-time steps.

\ms
Our approaches to {\sc AutLifting} each involve defining a group $\calL$ such that

\ms

\hspace*{1em}\parbox{4in}{
\begin{itemize}
\item $\Aut(G)_H \le \calL \le  \Sym(G)_{H,G/H}$.\\[-.45em]
\item It is ``easy'' to find $\,\Theta(\calL )\cap (\calA\times \calB).$\\[-.45em]
\item It is ``easy'' then to lift  to $\,\widehat{\calL}=\calL \cap \Theta^{-1}(\calA\times \calB).$\\[-.45em]
\item It is ``easy'' to find $\,\widehat{\calL} \cap \Aut(G)$.
\end{itemize}
}

\ms

\subsection{First choice of $\calL$}
\label{firstL}

We consider the following supergroup of $\Aut(G)_H$.

$$
\calL_1~=~
\{ \gamma\in \Sym(G)_{H,G/H} \,\mid\,
\forall g\in G, h\in H:~ (hg)^\gamma = h^\gamma g^\gamma\}.
$$

\ms
\subsubsection{Theory}

\ms

\begin{lemma}
\label{kerL1}
{\sl
{~}\\[-3ex]
\begin{enumerate}[\rm (i)]
\item
The kernel of $\,\restr{\Theta}{\calL_1}\,$ is isomorphic to the direct product
of $~|G/H| -1~$ copies of $\,H$.
\item
(Generators for) $\,\Ker(\restr{\Theta}{\calL_1})\,$ can be found in polynomial time.
\end{enumerate}
}
\end{lemma}
\goodbreak
\mni
{\sc Proof:}
Let $\gamma\in\Ker(\restr{\Theta}{\calL_1})$.   Then $\forall h\in H, \, h^\gamma = h$ and
$\forall x\in G,\, (Hx)^\gamma = Hx$.   
Consider the action induced by $\gamma$ on $Hg \ne H$.
We have $g^\gamma = gk$ for some $k\in H$ (since $gH=Hg$).   Then for any $hg\in Hg$,
$(hg)^\gamma = h^\gamma g^\gamma = hgk$, that is, $\forall x\in gH$, $x^\gamma = xk$.
Thus, the action of $\Ker(\restr{\Theta}{\calL_1})$ on each $Hg\ne H$
is the group of right multiplications by $H$, a group isomorphic to $H$.   Since the actions are independent
across the cosets,  (i) follows.

For (ii), we deal with each $Hg\ne H$ independently.  Focussing on a coset $X$,
we form for each $h\in H$, the permutation $\phi_{X,h}$ of $G$ that fixes all $x\not\in X$
and maps $x\mapsto xh$ for $x\in X$.     Then $\Ker(\restr{\Theta}{\calL_1})$
is generated by $\{\phi_{X,h} \mid h\in H, X\in G/H \mbox{ with } X\ne H  \}$.
Take the union of all these permutations for all $Hg\ne H$.
(A bit more economically, only use $h$ in a generating set for $H$.)
  \QED

\bni
\begin{lemma}
\label{L1lift}
{\sl $\Aut(G/H)\times \Aut(H) \le \Theta(\calL_1)$.
Furthermore, given $(\alpha,\beta)\in\calA\times\calB$,
$\calL_1 \cap \Theta^{-1}(\alpha,\beta)$ can be constructed
in polynomial time.
}
\end{lemma}

\mni
{\sc Proof:}  It suffices to show,
for any $\alpha\in \Aut(G/H),\,\beta\in\Aut(H)$, we can construct a {\it single\/}
$\gamma\in\calL_1$,
for which $\Theta(\gamma)=(\alpha,\beta)$ since, by Lemma~\ref{kerL1}  the coset
$\,\gamma \,\Ker(\restr{\Theta}{\calL_1})$ then comprises the set
of preimages  of $(\alpha,\beta)$ in $\calL_1$.

We construct $\gamma\in\calL_1$ as follows.    For $h\in H$,
define $h^\gamma := h^\beta$.  For each coset of $X$ of $H$ in $G$ with $X\ne H$,
define $\gamma$ on $X$ as follows:

\hspace*{1em}\parbox[t]{4in}{
\begin{enumerate}[1.]
\item
Fix  any $a\in X$ (so $X=Ha$).\\[-1.0em]
\item Fix {any} $~b\in  X^\alpha$. \\[-1.0em]
\item For all $h\in H$, define $(ha)^\gamma := h^\beta b$.
\end{enumerate}
}

\vspace*{-1em}
\QED

\bs\bs

Using Lemmas \ref{kerL1} and \ref{L1lift}, we conclude

\mni
\begin{proposition}
\label{L1find}
{\sl
{~}\\[-2ex]
\begin{enumerate}[\rm (i)]
\renewcommand{\theenumi}{\roman{enumi}}

\item
Given  $\,\calA\le \Aut(G/H)$ and $\calB \le \Aut(H)$,
$\widehat{\calL}_1:=\calL_1 \cap \Theta^{-1}(\calA\times\calB)$ can be constructed in polynomial
time.
\smallskip
\item
$\widehat{\calL}_1$ is an extension of $\,\calA\times\calB\,$ by a direct product of
copies of $H$.
\end{enumerate}
}
\end{proposition}

\mni
{\sc Proof:} For (i), $\widehat{\calL}_1$ is generated by the lifts of the generators of $\,\calA \times \calB\,$
together with generators of $\,\Ker(\restr{\Theta}{\calL_1})$. \QED

\ms

\subsubsection{Algorithm}
\label{Algorithm1}

~\\

\noindent
{\bf Step 1.}  $\widehat{\calL}_1:=\calL_1 \cap \Theta^{-1}(\calA\times\calB)$
 
\mni
{\sc Method:}
By Lemma~\ref{L1find}(i), this is in polynomial time for any $\calA, \calB$. \QED

\bni
{\bf Step 2.}
Find $~\{\gamma \in \widehat{\calL}_1 \mid \gamma\in\Aut(G)\}$.

\mni
{\sc Method:} This can be expressed as a set-stabilizer problem for the natural extension of
$\widehat{\calL}_1 \le \Sym(G)$
to an action on $G\times G\times G$.   The set to stabilize
is $\{(a,b,ab) \mid a,b\in G\}$.  \hspace*{2em}  \QED

\bs
\begin{remark}{\rm 
In an inductive approach to \CSAuto~for solvable groups the calls to \AutLift~result
in a solvable $\widehat{\calL}_1$, thus putting Step 2 in polynomial time.
(We provide this forecast in the expectation
 that some readers would like to finish the solvable case
as an exercise.)
A more restrictive $\calL$ will yield our main results for general groups,
thus making  this section superfluous.   Nevertheless,
we retain this discussion of $\calL_1$.     By switching back to $\calL_1$ in \S\ref{topdown}
whenever $H$ is solvable, we limit the requisite machinery to the early paper \cite{Luk82}.}
\end{remark}

\ms

\subsection{Second choice of $\calL$}
\label{secondL}

Consider now a more restricted supergroup of $\Aut(G)_H$.

$$
\calL_2~=~
\{ \gamma\in \Sym(G)_{H,G/H} \,\mid\,
\forall g\in G, h\in H,\, (hg)^\gamma = h^\gamma g^\gamma~\mbox{and}~  (gh)^\gamma = g^\gamma h^\gamma \}.
$$

\ssk

\subsubsection{Theory}
\label{theory}

The advantage of cutting down from $\calL_1$ to $\calL_2$  is that
$\Ker(\restr{\Theta}{{\calL}_2})$ is  abelian,
which plays a role in guaranteeing polynomial-time
set-stabilizers for 
the instances of $\widehat{\calL}_2$ that arise.  (Actually, we only need
that kernel to be solvable.)~
However, since
$\calA \times \calB$ may not be contained in
$\Theta(\calL_2)$, we will first have to determine the
liftable subgroup.
We are guided in this by some properties of $\Theta(\calL_2)$.

\mni
\begin{lemma}
\label{gammaconj}
{\sl
Let $\gamma\in  \calL_2$.  Then for $g\in G$, $h\in H$,
$(g^{-1}hg)^\gamma = (g^\gamma)^{-1}h^\gamma g^\gamma$.}
\end{lemma}

\mni
{\sc Proof:}
For $g\in G$, $h\in H$, $\gamma\in\calL_2$,
$$
g^\gamma(g^{-1}hg)^\gamma =  (gg^{-1}hg)^\gamma
=  (hg)^\gamma
= h^\gamma g^\gamma,
$$
the first and third equalities using properties of $\calL_2$.   \QED

\bni
\begin{lemma}
\label{kerL2}
{\sl
{~}\\[-3ex]
\begin{enumerate}[\rm (i)]
\item
The kernel of $\restr{\Theta}{\calL_2}$ is isomorphic to the direct product
of $~|G/H| -1~$ copies of $\Z(H)$ (the center of $H$).

\ssk
\item
(Generators for) $\Ker(\restr{\Theta}{\calL_2})$ can be found in polynomial time.
\end{enumerate}
}
\end{lemma}

\mni
{\sc Proof:}
Let $\gamma\in\Ker(\restr{\Theta}{\calL_2})$.
In particular, $h^\gamma = h$ for $h\in H$.    
Consider $Hg\ne H$.
As in the proof of Lemma~\ref{kerL1}, there is some $k\in H$
such that the action of $\gamma$ on $Hg=gH$ is right-multiplication by $k$.
It suffices then to show $k\in \Z(H)$.  Using Lemma~\ref{gammaconj}, 
for any $h\in H$,
$$g^{-1}hg = (g^{-1}hg)^\gamma = (g^\gamma)^{-1}h^\gamma g^\gamma 
= (gk)^{-1}h gk = k^{-1}(g^{-1}hg)k.$$
The normality of $\Z(H)$ in $G$ then implies $k\in \Z(H)$.

For (ii), we proceed as in Lemma~\ref{kerL1}(ii) but restrict the choice of maps $x\mapsto xh$ to
$h\in \Z(H)$  (or just to a generating set of $\Z(H)$).
   \QED

\ms

\goodbreak

\skni

\goodbreak

\bs
The next two lemmas give necessary
conditions on $(\alpha,\beta) \in \Aut(G/H) \times \Aut(H)$ for it
to be liftable to
$\calL_2$.  

Recall that $\C$ denotes
the centralizer of
$H$ in $G$, and $\inn{x}$ denotes the restriction to $H$ of the inner automorphism
corresponding to $x$.

\bni
\begin{lemma}
\label{L2property}

{\sl
Let $\gamma\in  \calL_2$.   Suppose that
$\Theta(\gamma) = (\alpha,\beta) \in \Aut(G/H) \times \Aut(H)$.   Then
\begin{enumerate}[\rm (i)]
\renewcommand{\theenumi}{\roman{enumi}}
\item
$\alpha$ normalizes ${\CH}/{H}$, and therefore induces an automorphism of $G/\CH$.
\medskip
\item
$\forall g\in G:~\beta^{-1} \,\InnH{\CH g} \, \beta =\InnH{(\CH g)^\alpha}.$
\end{enumerate}
}
\end{lemma}

\mni
{\sc Proof:}
Lemma~\ref{gammaconj} implies that $\C^\gamma = \C$, so that 
$(H\C)^\gamma = H^\gamma \C^\gamma$, proving (i).\\
For (ii),  first note that $\forall g\in G, h\in H$,
$$ h^{\beta^{-1} \InnH{g} \beta}  = (g^{-1} h^{\beta^{-1}} g)^\beta
 = (g^\gamma)^{-1} h g^\gamma = h^{\InnH{g^\gamma}},
$$
the second equality following  from Lemma~\ref{gammaconj}.
In other words,
$$ \forall g\in G,~
\beta^{-1}\InnH{g} \beta = \InnH{g^\gamma}.$$

\vspace*{-3ex}
\QED

\bni
\begin{lemma}
\label{L2property2}
{\sl Suppose $(\alpha,\beta) \in \Aut(G/H) \times \Aut(H)$ satisfies {\rm (i),(ii)} in
Lemma~\ref{L2property}.
Then
\begin{equation}
\label{cosetconj}
\forall g\in G:~\beta^{-1} \left(\InnH{H g} \right) \beta =\InnH{(H g)^\alpha}.
\end{equation}
}
\end{lemma}

\mni
{\sc Proof:}
For any $A\subseteq G$,
$\InnH{A \C}= \InnH{A}$.   Also, by (i),
$\alpha$ permutes the cosets of $H\C/H$ in $G/H$ so that
$(H\C g)^\alpha = \C(Hg)^\alpha$.
\QED

\bni
\begin{remark}{\rm
This organization may seem convoluted seeing that
equation~(\ref{cosetconj})
could also be viewed as a direct consequence of
Lemma~\ref{gammaconj}.    Our motive is that, in the process of cutting down to
``liftable''
$(\alpha,\beta)$,
our {\it algorithmic\/}
route runs through (\ref{cosetconj}) after first forcing (i),(ii) of Lemma~\ref{L2property}.}
\end{remark}

\goodbreak

\bni
\begin{lemma}
\label{L2find}
{\sl
Suppose $(\alpha,\beta) \in \Aut(G/H) \times \Aut(H)$ satisfies property~{\rm (\ref{cosetconj})}. \\
Then
$\,\Theta^{-1}(\alpha,\beta)\cap\calL_2\,$ is nonempty and can be found in polynomial
time}.
\end{lemma}

\mni
{\sc Method:}
Construct $\gamma\in\Theta^{-1}(\alpha,\beta)$ as follows.

For $h\in H$, $h^\gamma := h^\beta$.

To define $\gamma$ on a coset $X$ of $H$ with $X\ne H$.   Fix $a \in X$.
Then, by (1), there is some $b\in X^\alpha$ such that
$~
\beta^{-1}\InnH{a}\beta =  \InnH{b}.
~$
Fix any such $b$.  For any $g\in X$, say $g=ka$ with $k\in H$,  
set $g^\gamma := k^\beta b$.
It follows easily that
\begin{equation}
\forall h\in H, ~  (hg)^\gamma = h^\gamma g^\gamma
\end{equation}
Also
\begin{equation}
 \beta^{-1}\InnH{g}\beta =  \InnH{g^\gamma}
\end{equation}

\skni
since~ 
$\beta^{-1} \InnH{ka} \beta =
\beta^{-1} \InnH{k} \beta\, \beta^{-1} \InnH{a} \beta =
\InnH{k^\beta} \InnH{b} = \InnH{k^\beta b}.
$

\ms
Having established (2) and (3) for $g$ in each coset $X$, they hold for
all $g\in G$.

\ms
Then, for all $h\in H, g\in G$,
$$(gh)^\gamma = (h^{\inn{g}^{-1}}g)^\gamma = h^{{\inn{g}^{-1}}\beta}g^\gamma
= h^{\beta (\beta^{-1}  \inn{g} \beta)^{-1}}g^\gamma =
h^{\beta \, \inn{g^{\gamma}}^{-1} }g^\gamma = g^\gamma h^\gamma.
$$
We conclude that $\gamma\in \calL_2$.  

\ms 
The complete set of preimages of
 $(\alpha,\beta)$ is
 $\gamma\, \Ker(\restr{\Theta}{\calL_2})$, for which we apply
Lemma~\ref{kerL2}(ii).  \mbox{\hspace*{3em}}
\QED

\bs
To summarize the main points of this subsection, we have the following
consequence of
Lemmas \ref{kerL2}, \ref{L2property}, \ref{L2find}.

\goodbreak
\bni
\begin{proposition}
\label{L2summary}
{\sl Suppose  $\calM\le \Aut(G/H) \times \Aut(H)$ such that
  property~{\rm (\ref{cosetconj})} is satisfied
by all $(\alpha,\beta)\in \calM$.  Then
\begin{enumerate}[\rm (i)]
\item
$\calM \le \Theta({\calL}_2)$
\ssk
\item
$\calL_2\cap\Theta^{-1}(\calM)$ is an extension of $\calM$ by an abelian group.
\ssk
\item
Given $\calM$, $\calL_2\cap\Theta^{-1}(\calM)$ can be constructed in polynomial time.

\end{enumerate}
}
\end{proposition}

\mni
{\sc Proof:} For (iii), $\calL_2\cap\Theta^{-1}(\calM)$ is generated by the lifts of the generators of $\,\calM\,$
together with generators of $\,\Ker(\restr{\Theta}{\calL_2})$. \QED

\ms

\subsubsection{Algorithm}
\label{algorithm}

Steps 1-3 cut $\calA\times\calB$ to the subgroup $\calM$ consisting of elements that
can be lifted to $\calL_2$.

\bni
{\bf Step 1.}
$\calA\,:=\, \nrm{\calA}{{\CH}/{H}}$.

\mni
{\sc Method:}
Viewing $\calA\le \Sym(G/H)$,
this can be viewed as stabilizing the subset $\CH/{H}$
of the polynomial-size domain $G/H$.    \QED

\ms
By Lemma~\ref{L2property}(i), Step 1 does not affect
$~\Theta(\calL_2) \cap (\calA\times\calB)$.

\bni
{\bf Step 2.} $\calB \,:=\,  \nrm{\calB}{\InnH{G}}$.

\mni
{\sc Method:}
Here we consider $\calB \subseteq \Sym(H)$  acting on
$\Aut(H)\subseteq\Sym(H)$ via conjugation.  So, at first glance, this
appears to  be a normalizer problem for permutation groups.
However, the specific reductions of
our main problems to \AutLift~will reveal that required
instances of Step 2 can be handled via set-stabilizers.
\QED

\ms
By Lemma~\ref{L2property}(ii), Step 2 does not affect
$~\Theta(\calL_2) \cap (\calA\times\calB)$.   

\bs
Step 2 does not yet accomplish the
compatibility condition on $(\alpha,\beta)$ expressed in property (1)
and required in Lemma~\ref{L2find},
but it establishes a structure for getting there.

\bs

We now have that $\calA \le {\Aut(G/H)}_{\CH/H}$, so there is an induced action
$$\fra: \calA \rightarrow \Aut\left(\frac{G}{\CH}\right).
$$

\nin
We also now have $\calB$ acting on $\InnH{G}$ and, since $\calB\le\Aut(H)$
also normalizes $\InnH{H}$, there is an induced action
$$\frb:  \calB \rightarrow \Aut\left(   \frac{\InnH{G}}{\InnH{H}}    \right)
$$

\nin
But there is a natural  identification
$$
\frac{G}{\CH} \,\,\,{\simeq}\,\,\,
\frac{\InnH{G}}{\InnH{H}}.$$

\bni
(For all $g\in G$, $g\CH \leftrightarrow\InnH{g}\InnH{H}$.)$~~$
Via this identification, property (1)
is expressible in the form

$$
\fra(\alpha)=\frb(\beta).
$$

\mni
This motivates
\goodbreak
\mni
{\bf Step 3.}  $\calM  \,:=\, \{(\alpha,\beta)\in\calA\times \calB \mid \fra(\alpha)=\frb(\beta)\}$.

\mni
{\sc Method:}
By the above, this is a permutation-group intersection problem, solvable for example as a
set-stabilizer problem:    consider  $\fra(\calA)\times \frb(\calB)$ acting on $\,\Omega\times \Omega\,$
where  $\,\Omega = G/\CH$;  the object then is
to stabilize the diagonal $\{(\omega,\omega) \mid \omega\in \Omega\}$.   \QED

\bs

By Lemmas~\ref{L2property},\ref{L2property2} and Proposition~\ref{L2summary}(i),
$~\Theta(\calL_2) \cap (\calA\times\calB)\,=\, \calM$.
Hence, $\calL_2\cap\Theta^{-1}{(\calA\times\calB)} \,=\, \calL_2\cap\Theta^{-1}(\calM)$.
So the next step is

\bni
{\bf Step 4.}
$\widehat{\calL}_2  \,:=\ \calL_2\cap\Theta^{-1}(\calM)$.

\mni
{\sc Method:}
This is in polynomial time by Proposition~\ref{L2summary}(iii).    \QED

\bs
The final step is

\bni
{\bf Step 5.}
Find $~\{\gamma \in \widehat{\calL}_2 \mid \gamma\in\Aut(G)\}$.

\mni
{\sc Method:}
As in \S\ref{Algorithm1}, 
this can be expressed as a set-stabilizer problem for the  action of
$\widehat{\calL}_2$ on $G\times G\times G$.    \QED

\bs\ms
Thus the computational complexity of our use of
\AutLift~rests on properties of $\calA$ and $\calB$ that will enable efficient
routines for Steps 1,2,3,5.    The properties are
described in \S\ref{nice}.

\bs

\section{Nice groups}
\label{nice}

\subsection{A reminder on polynomial-time set stabilizers}
\label{setstab}

\ms

In \cite{Luk82},  the author proposed
an algorithm for finding $G_\Delta$ where $G\le \Sym(\Omega)$
and $\Delta\subseteq \Omega$.    For the reader's convenience in what follows, we offer a
brief description.\footnote{See also  \cite{LM11} for an extended discussion of the
divide-and-conquer paradigm for this and other applications.}

\ms

To accommodate a recursion,
the problem is generalized to cosets.

\bs
\ms

\begin{minipage}{\textwidth}
\begin{quote}
For a $G$-stable subset
$\,\Pi \subseteq \Omega\,$ and  a coset $\,X=Ga\,$ of $\,G$, define
$$X_{\Delta \mid \Pi} = \{ x\in X \mid (\Delta\cap\Pi)^x = \Delta\cap\Pi^a\}.
$$
So  $\,X_{\Delta \mid \Pi}\,$ is either $\emptyset$ or a right coset of $\,G_{\Delta\cap\Pi}$.
If $\,\Pi = \Pi_1\dot\cup\Pi_2\,$ for $G$-stable $\,\Pi_i$
$$X_{\Delta \mid \Pi} ~:=~ (X_{\Delta \mid \Pi_1})_{\Delta\mid \Pi_2}
$$
If $G$ acts transitively on $\Pi$ and $|\Pi| >1$, find a minimal
decomposition $\,\Pi = \Pi_1 \dot\cup \cdots \dot\cup \Pi_m\,$ into blocks
of imprimitivity and decompose $\,G= \bigcup_{1\le i\le |G/H|} Ht_i\,$ where $H$
is the kernel of the action of $G$ on $\,\{\Pi_i\}_{1\le i\le m}$.
$$X_{\Delta \mid \Pi} ~:=~ \bigcup_{1\le i\le |G/H|}  (Ht_ia)_{\Delta \mid \Pi}.
$$
If $\,\Pi =\{\pi\}\,$ then\\
\centerline{if $\,|\Delta \cap \{\pi,\pi^a\}|=1\,$ then $\,X_{\Delta \mid \Pi} := \emptyset\,$
else $\,X_{\Delta \mid \Pi} := X$.}
\end{quote}
\end{minipage}

\bs
\bni
The computation of $\,G_\Delta\,$ starts with $\,\Pi:=\Omega\,$ and $\,a:=1$.
The key to the complexity of the method lies in the sizes of the primitive
groups arising in the action on the $m$ blocks.     The algorithm will run in polynomial time for an hereditary
class of groups if such induced subgroups of $S_m$
have order $O(m^{\rm constant})$.

\ms

Notice that the ``set-stabilizer'' algorithm  actually dealt with cosets.    Thus,  we can find set-stabilizers for a
group $A$ that has a small-index subgroup $B$ in a good class:  we start by
breaking $A$ into cosets of $B$.
We will find ourselves in exactly
that situation in \S\ref{almostsol}.

\def\nega{{\mbox{\tiny -1}}}

\bni
\begin{remark}
\label{cosetstab}
{\rm
 If we were only given a black-box for set-stabilizers in groups with bounded non-cyclic composition factors, i.e.,
without knowing the algorithm, we could still use that for the coset problem.      It is useful
for another purpose in \S\ref{bottomup} to express that as a ``set-transporter'' problem;
namely, given $\Delta,\Lambda \subseteq \Omega$, find
$$G_{\Delta \mapsto \Lambda } = \{g\in G \mid  \Delta^g = \Lambda\}.$$
For this, consider
the natural action of $\widetilde{G}= G\wr C_2$
on $\Omega\,\dot\cup\, \Omega'$,  where  $\Omega'$ is a copy of $\Omega$.
The desired transporter can be deduced from
$\widetilde{G}_{\Delta \,\dot\cup \,\Lambda'}$ where $\Lambda'$ is the corresponding
copy of $\Lambda$.    In particular, $(Ga)_\Delta = 
G_{\Delta\mapsto \Delta^{a^{\nega}} }\,a$.
}
\end{remark}

\ms

\subsection{The groups that turn up}
\label{almostsol}

In effect, we will only rely on the polynomial boundedness
of primitive solvable groups.   However, we deal with groups
that are not quite solvable.

\mni
{\bf Notation.} For a group $X$, $\Sol(X)$ is the solvable radical of $X$, i.e., the maximum normal
solvable subgroup.

\mni
{\bf Definition.}  Let $X\le \Sym(\Omega)$.  We call $X$ {\it \nice\/}$\,$\footnote{
The author regrets using a term that has had other meanings.   However, the usage
of ``\nice'' is already inconsistent in the literature and he could not think of an  
{\it unused\/} term with
similar connotation.}$\,$   if
$\,|X:\Sol(X)| \le |\Omega|^2$.

\bs
\Nice~groups still enable the 
polynomial-time divide-and-conquer paradigm.   For example,

\mni
\begin{lemma}
\label{nicestab}
{\sl Given an \nice~$G\le\Sym(\Omega)$ and $\Delta\subseteq\Omega$, 
$G_\Delta$ can be found in polynomial time.}
\end{lemma}

\mni
{\sc Proof:}
By decomposing $\,G\,$ into cosets of $\,\Sol(G)\,$, we reduce to
 $|\Omega|^2$ problems involving
solvable groups.   \QED

\bni
\begin{remark}{\rm
Our breakdown to cosets of a nice group is reminiscent of the first use of the set-stabilizer method.
When \cite{Luk82} was written,
primitive groups in the relevant class, specifically the class of groups with bounded composition factors, were not known to 
be polynomially bounded.
This difficulty was overcome by partitioning the primitive
group into cosets of a small-index $p$-subgroup.   That additional
complication soon became unnecessary as polynomial bounds were
established 
 first for primitive solvable
groups (independently by P\'{a}lfy \cite{Pa82} and Wolf \cite{Wo82}), and ultimately for 
primitive groups in a
class that even includes groups with bounded {\it non-cyclic\/} composition factors  
(Babai {\sl et al.}~{\cite{BCP82}).      We
point, however, to 
a subtle difference between what happens in our current passage
to cosets of a nice group and the earlier use of that trick.
In  \cite{Luk82}, after passing to cosets of a  $p$-group acting on the blocks, the
divide-and-conquer narrows our window to a single, stabilized block.    Since the group acting within the block is not necessarily
a $p$-group, we may again arrive at a primitive group that requires the cosets-of-$p$-group
decomposition.
By contrast, in the present
case there is a {\it single\/} such decomposition in the lifetime of the set-stabilizer process, i.e.,
we only visit solvable groups thereafter. }
}\end{remark}

\mni
\begin{remark}{\rm
Lemma~\ref{nicestab} is a weak consequence of the 
discussion in \S\ref{setstab}.    
The good subgroup can be of any polynomially-bounded index,
need not be normal,
and need only be in the broader class
described in \cite{BCP82}.   
However, ``\nice'' is a good fit for our situation because we next see that the
property arises so conveniently.
Furthermore, in our present application, one can keep track of small-index normal
solvable subgroups
as we cut down the group or take preimages, so there
is no need even to implement a radical-finder.   Nevertheless,
we do note that radicals of
permutation groups can be found in polynomial-time 
 (\cite[Lect.~6]{Luk93}, \cite[\S6.3.1]{Se03}).
}\end{remark}

\ms
The relevance to our problem is seen in 

\skni\goodbreak
\begin{lemma}
\label{L1L2sol}
{\sl
With reference to the groups in \S2,
suppose $\calA\le \Sym(G/H)$ and $\calB\le \Sym(H)$ are \nice.
\begin{enumerate}[\rm (i)]
\item
If $H$ is solvable, then $\widehat{\calL}_1\le\Sym(G)$ is \nice.
\ssk
\item
$\widehat{\calL}_2\le\Sym(G)$ is \nice.
\end{enumerate} }
\end{lemma}
\goodbreak
\mni
{\sc Proof:}
We have $|\calA\times\calB:\Sol(\calA\times\calB)| = 
| \calA:\Sol(\calA)| \cdot |\calB:\Sol(\calB) | \le |G/H|^2|H|^2 = |G|^2$.
Thus, (i) follows from Proposition~\ref{L1find}(ii).

\ssk
Using $\calM$ as in \S\ref{algorithm} (Steps 3,4), $\,\calM\le \calA\times\calB$ implies
$|\calM : \Sol(\calM)| \le |\calA\times\calB:\Sol(\calA\times\calB)| \le |G|^2$.
Thus (ii) follows from Proposition~\ref{L2summary}(ii).    \QED

\bs
Note, as the methods for \AutLift~are to be used repeatedly, the \niceness~of
$\,\widehat{\calL}_i\,$  implies that of the subgroup $\,\widehat{\calL}_i \cap \Aut(G)$.

For a base case, we need a consequence of the Classification of Finite Simple Groups.
Using the fact that any simple group $T$ can be generated by two elements (\cite{AB}),
we immediately have $|\Aut(T)| \le |T|^2$.  Hence,

\mni
\begin{lemma}
\label{simplenice}
{\sl If $T$ is simple then $\Aut(T)$, viewed as a subgroup of $\Sym(T)$, is
\nice.}
\end{lemma}

\mni
In fact, it is well known that the Classification 
 yields a stronger bound of the form $|\Aut(T)| = O( |T| \log |T|)$ (e.g., see \cite{ATLAS}).
However, the square bound in ``\nice'' is easier to maintain
through our process.

\goodbreak

\section{Automorphisms stabilizing  a composition series}
\label{auto}

\ms
\begin{sloppy}
Applying the machinery of \S\S\ref{key}-\ref{nice}, we offer two
polynomial-time Turing reductions of \CSAuto~to polynomial-time instances of \AutLift.
Moreover, we show that the
deepest tool needed in  either implementation is 
set-stabilizer for solvable permutation groups.
Aside from reducing to the most basic tool, this extra effort will be useful
in a subsequent development of canonical forms. (See also \cite{BaLu83}
for an indication of how the divide-and-conquer method for set-stabilizer
translates to canonical set placement.)
\end{sloppy}

\subsection{Bottom-up on given series}
\label{bottomup}

\ms
We go from a solution for $G_i$ to a solution for $G_{i-1}$.
The group on top, $G_{i-1}/G_i$, is simple and so
one can enumerate $\Aut(G_{i-1}/G_i)$.

\mni
\begin{remark}{\rm Even for a simple {\it permutation group\/}  $T$ given
by generators, one can produce
{generators for} $\Aut(T)$ (which is all that is necessary for some of our subproblems).
This follows from Kantor's demonstration \cite{Kan85}
that one can obtain the ``natural'' representation
of $T$.
}\end{remark}

\mni
\begin{remark}{\rm
Before proceeding further,  
we can claim  to have already established a poly\-no\-mial-time solution for this use of
\AutLift.  That is, by virtue of almost-solvability (\S\ref{nice}), there are citable polynomial-time methods for
Steps 1,2,3,5~in \S\ref{secondL}.  Thus,
a message of  this subsection is that we do not need the
full power of available polynomial-time tools.   That leads to speculation
that there are  more general problems to solve.  
}\end{remark}

\ms
Let us consider the steps of the algorithm in \S\ref{algorithm}.
Since there are more interesting things to say about Step 2, we will postpone that discussion and
dispense with the other steps.

\bni
{\bf Step 1:}   
Since we always arrive at \AutLift~with $G/H$ simple,
either $\,\CH=G\,$ or $\,\CH= H$.   In either case, $\calA$ already normalizes $\CH/H$ so there is nothing
more to do.

\bni
{\bf Step 3:}
This is a group intersection.  So it  would be in polynomial time if just one group is
\nice~\cite[ \S4.2]{Luk82} and, as indicated,
just a polynomial-time {\it set-stabilizer\/} if both groups are \nice, as is the case here.
However, in this situation,
none of the machinery is even needed because
$\calA$ is listable.    Note also that this  is trivial  when $\,\CH = G\,$  (which means $\calM = \calA\times\calB$). 

\bni
{\bf Step 5:}
As indicated, this is a set-stabilizer for a group that we now know to be \nice.

\bs Now for {\bf Step 2:} 

\mni
Three approaches will be indicated, winding up with just set-stabilizers.   
We state these for a
broader class than \nice~groups.    

For an integer constant $d > 0$,
let $\,\Gamma_d\,$ denote the class of finite 
groups all of whose nonabelian composition factors lie in
$\,S_d$.  Also, bear in mind that the polynomial timing
for $\,\Gamma_d\,$ groups immediately  extends to situations
where we are in possession of a $\,\Gamma_d\,$
subgroup of polynomial index (see \S\ref{nice}.)~
In particular, these methods apply to
\nice~groups.


\PROBLEM{
\medskip
{\sc Problem (I)}\\[1ex]
\hspace*{1em} {\sc Given:} $X,Y\le \Sym(\Omega)$ with $X\in \Gamma_d$.\\[1ex]
\hspace*{1em} {\sc Find:}
$X_Y$.    \hfill (Where $X$ is acting on $\Sym(\Omega)$ via conjugation.)
\\[-.75ex]
}

\mni
{\it Method.}  
This was shown to be in polynomial-time by Luks and Miyazaki \cite{LM11}.
\footnote{While $\Gamma_d$ is slightly smaller
than the class available just for set-stabilization \cite{BCP82}, the restriction
is still needed for this method.}     \QED

\bs 

In our situation, we are trying to normalize a {\it polynomial-size\/} $Y$ and there
is a more elementary approach to this special situation.

\PROBLEM{
\medskip
{\sc Problem (II)}\\[1ex]
\hspace*{1em} {\sc Given:} $X,Y\le \Sym(\Omega)$ with $X\in \Gamma_d$ and
$|Y|= {\rm O}(|\Omega^{\rm const}|)$.\\[1ex]
\hspace*{1em} {\sc Find:}
$X_Y$
\\[-.75ex]
}

\mni
{\sc Method:}~
Let $\Sym(\Omega)$ act on
$\Omega\times\Omega$ diagonally.    Also, for $s\in\Sym(\Omega)$, let
$$\Delta_s \,:=\, \{(\omega,\omega^s) \mid \omega\in\Omega\}.
$$
Then for $y,x\in\Sym(\Omega)$,
$$\Delta_y^x\,=\, \Delta_{x^{-1} y x}.
$$
Thus, $x$ normalizes $Y$ iff $x$ stabilizes the collection
$\{\Delta_y\}_{y\in Y}$.   So, finding $X_Y$ is a matter of finding the
subgroup of $X$ inducing automorphisms of a hypergraph.   Miller \cite{Mil83b}
has shown that  to be in polynomial time for $X\in \Gamma_d$.   \QED

\bs

Our situation is even more special.

\PROBLEM{
\medskip
{\sc Problem (III)}\\[1ex]
\hspace*{1em} {\sc Given:} $X,Y\le \Sym(\Omega)$ with $X\in \Gamma_d$ and
$|Y|= {\rm O}(|\Omega|^{\rm const})$;\\
\hspace*{4.7em} $K< Y$ with $K\normal \langle X,Y\rangle$ and 
we are able to list $\Aut(Y/K)$.
\\[1ex]
\hspace*{1em} {\sc Find:}
$X_Y$.
\\[-.75ex]
}

\mni
{\sc Method:}~
Note that $X_Y=X_{Y/K}$.   
For each $\sigma\in\Aut(Y/K)$, we find those  $x\in X$ such that conjugation
by $x$ induces $\sigma$.   This will be case  iff

\ms

\centerline{$\forall y\in Y:  x^{-1}y x \,\in \, (yK)^\sigma $}

\noindent
which is the case iff

\ms

\centerline{$\forall y\in Y, \exists z \in(yK)^\sigma:  \Delta_y^x = \Delta_{z}$}

\mni
(See Remark~\ref{cosetstab} for viewing these
``set-transporters'' as set-stabilizers.)~ It is feasible to run through all $y$ and then all $z$.   \QED

\goodbreak
\bs

To apply the {\sc Problem~(III)} method in
our situation, $Y=\InnH{G}$, $K=\InnH{H}$.    
Furthermore, we are only concerned with the case $\CH=H$, else
$Y=\InnH{G}=\InnH{H}$ 
which is already
normalized by $\calB$.   Thus $Y/K$ is simple.   
So we can not only list
$\Aut(Y/K)$ but we  can even shorten the process by checking that
$x$ conjugates correctly on just  $y_1, y_2\in Y$, where $y_1K, y_2K$ generate
$Y/K$.

\ms
A further savings on the number of set-stabilizer calls can be realized by using
a quotient-group method of Kantor and Luks  \cite[problem {P7}(ii)]{KL90}.
Finding the $x\in X$ such that $(yK)^x = (yK)^\sigma$ can be accomplished 
with a single set-stabilizer.

\subsection{Top-down on a refinement of a characteristic series}
\label{topdown}

\mni
Recall that a subgroup $H$ of a group $G$ is called {\it characteristic\/} if it is invariant
under $\Aut(G)$.   A group with no proper characteristic subgroups is called 
{\it characteristically simple} and is necessarily a product of isomorphic simple groups.
A {\it characteristic series\/} in $G$ is a chain
$G=K_0 \vartriangleright K_1  \vartriangleright K_2  \vartriangleright \cdots \vartriangleright K_r =\1$
for which each $K_i$ is characteristic in $G$.   

A characteristic series can be constructed
with characteristically simple quotients\linebreak {${K_i/K_{i+1}}\,$}
even if $G$ is a permutation group given by generators; see, e.g., \cite{KL90}.   For groups
given by a Cayley table, the construction is elementary:  find the minimal normal subgroups
of $G$ by considering the normal closures of all elements; these will each be
characteristically simple, so select those whose simple factors are of designated type
and let them generate the subgroup $K$.   Continue the process with $G/K$, etc.

\begin{lemma}
\label{CharSer}
{\sl 
Without loss of generality, we may assume that the series $G_0,G_1,G_2,\ldots,G_m$
in \CSAuto~has a characteristic subseries
$$G=K_0 \vartriangleright K_1  \vartriangleright K_2  \vartriangleright \cdots \vartriangleright K_r =\1$$
where each $K_i/K_{i+1}$ is characteristically simple.\footnote{
The Wagner-Rosenbaum composition series are already of the special type.}}
\end{lemma}

\noindent
{\sc Proof:}
We are given a composition series
$$G=G_0 \vartriangleright G_1  \vartriangleright G_2  \vartriangleright \cdots \vartriangleright G_m =\1.
$$
As indicated above, we construct  a {\it characteristic\/} series
$G=K_0 , K_1 , K_2  , \ldots, K_r =\1$
 with characteristically simple quotients $K_i/K_{i+1}$.  
Refine the series between each $K_i$ and $K_{i+1}$  by inserting
 $$K_i =(K_i\cap G_0)K_{i+1} \trianglerighteq (K_i\cap G_1)K_{i+1}  
 \trianglerighteq (K_i\cap G_2)K_{i+1}  \trianglerighteq \cdots \trianglerighteq (K_i\cap G_m)K_{i+1} =K_{i+1}$$
 and eliminate duplicates.
 We again have a composition series and 
automorphisms stabilizing the original series will stabilize this new one.     Having computed
 the automorphisms stabilizing the new series, we have an \nice~group, so
cutting down the result to
 stabilize the original series is done with set stabilizers.
\QED

\bni
{\it Method for Theorem~\ref{mainprobauto}.}

Having reset the series as in Lemma~\ref{CharSer},
successively compute the appropriate subgroup of $\Aut(K_0/K_i)$, where $(K_i)_i$ is
the embedded normal series as in Lemma~\ref{CharSer}.


In the base case $K_0/K_0$ is trivial.

For the inductive step, the call to
{\sc AutLifting} involves $K_0/K_{i+1}\,$  and its normal subgroup $K_i/K_{i+1}$.
We arrive with the inductive input $\calA  \le \Aut(K_0/K_{i})$.
The group $\calB$ should consist of
the automorphisms of the semisimple group $H := K_i/K_{i+1}$ that fix 
the composition
series induced on this section.
If $H$ is nonabelian, $\calB$ is the direct product of the
automorphism groups of the simple factors.   If $H$ is a product
of  cyclic groups of prime order $p$, $\calB$
can be viewed as the upper triangular matrices over ${\rm GF}(p)$.


Using the $\calL_2$ method at each stage, 
the procedure is in polynomial time for {{\it all groups}, but that seemed
to require hypergraph stabilizer in Step 2.
So let us revisit that step.

Suppose  $H$ is nonabelian.   Then $\calB$ (which now fixes
the simple factors) is of polynomial size, e.g.  $|\calB|\le |H|^2$ in which
case Step 2 can be carried out
by testing each element of $\calB$.  This is not the case if $H$ is abelian, but
then we simply revert to the 
$\calL_1$ method of \S\ref{firstL} for this round. \QED

\ms



\section{Isomorphism matching fixed composition series}
\label{Iso}

\ms
We prove Theorem~\ref{mainprob} via
a familiar technique
in isomorphism studies, applying the automorphism-group result to finding isomorphisms.
For example, an algorithm for finding automorphism groups of graphs will find the isomorphisms
between two connected graphs $X_1,X_2$ by finding the automorphism group of the disjoint
union $X_1\dot\cup X_2$.       An analogous construction works here.

\ms
We are given composition series
$$G_1=G_{1,0} \vartriangleright G_{1,1}  \vartriangleright G_{1,2}  \vartriangleright \cdots \vartriangleright G_{1,m}  =\1$$
$$G_2=G_{2,0} \vartriangleright G_{2,1}  \vartriangleright G_{2,2}  \vartriangleright \cdots \vartriangleright G_{2,m}=\1$$
Form the single subnormal series
$$G_1\!\times\!G_2=G_{1,0}\!\times\! G_{2,0}\vartriangleright G_{1,1}\!\times\! G_{2,1}  \vartriangleright 
G_{1,2} \!\times\! G_{2,2} \vartriangleright \cdots \vartriangleright G_{1,m}\!\times\! G_{2,m}  =\1\times\1.$$
We directly accommodate  the setup of \S\ref{bottomup}~to this situation.  (The
method of \S\ref{topdown} could be adapted as well.)~
We now want the automorphisms of $G_1\!\times\!G_2$ that not only fix the terms in this series but, in doing so,
fix or switch the factors.   If, for any $i$, we find that there are no relevant automorphisms of
$G_{1,i}\!\times\! G_{2,i}$ that switch factors, we exit with a negative
response to \CSIso. 

Calls to \AutLift~will now
have $G/H$ as the product of two isomorphic simple groups.   For $\calA$ we take the automorphisms
that fix or switch the factors.   (That would always be the case if the simple groups are nonabelian.)

The rest of the discussion, including the various methods for Step 2, proceed as before.

\begin{remark}
{\rm A  trivial technicality.
Our weak version
of Lemma~\ref{simplenice} does not quite say 
$\Aut({T\times{}T})\allowbreak\le\Sym (T\times T)$ is \nice~for
simple nonabelian $T$ since $|\Aut(T\times T)| = 2 |\Aut(T)|^2$.
As already noted, Lemma~\ref{simplenice} could be strengthened,
but it is clear that the theory is unaffected by an extra factor of 2.  
}
\end{remark}

\bs\goodbreak

\centerline{\sc acknowledgements}

\bs
The author is delighted to acknowledge communications with Laci Babai and
James Wilson that stimulated the development of these ideas
and improved their exposition.

\bs\ms

\def\bysame{\underline{\hspace*{4em}}}

\bibliographystyle{amsplain}

\goodbreak
\end{document}